# Comparison of tRNA and rRNA Phylogenies in Proteobacteria: Implications for the Frequency of Horizontal Gene Transfer

Bin Tang, Philippe Boisvert and Paul G. Higgs

Department of Physics and Astronomy,

McMaster University, Hamilton, Ontario L8S 4M1, Canada.

higgsp@mcmaster.ca Tel (905) 525 9140 ext 26870 Fax (905) 546 1252

**Abstract**

The current picture of bacterial evolution is based largely on studies of 16S rRNA. However, this is just one gene. It is known that horizontal gene transfer can occur between bacterial species, although the frequency and implications of this are not fully understood. If horizontal transfer were frequent, there would be no single evolutionary tree for bacteria because each gene would follow a different tree. We carried out phylogenetic analyses of rRNA and tRNA genes from Proteobacteria (a diverse group for which many complete genome sequences are available) using RNA-specific phylogenetic methods that account for the conservation of the secondary structure. We compared trees for 16S rRNA and 23S rRNA with those derived from concatenated alignments of 29 tRNA genes that are found in all the genomes studied. The tRNA genes are scattered throughout the genomes, and would not follow the same evolutionary history if horizontal transfer were frequent. Nevertheless, the tRNA tree is consistent with the rRNA tree in most respects. Minor differences can almost all be attributed to uncertainty or unreliability of the phylogenetic method. We therefore conclude that tRNA genes give a coherent picture of the phylogeny of the organisms, and that horizontal transfer of tRNAs is too rare to obscure the signal of the organismal tree. Some tRNA genes are not present in all the genomes. We discuss possible explanations for the observed patterns of presence and absence of genes: these involve gene deletion, gene duplication, and mutations in the tRNA anticodons.

**Introduction**

The 16S rRNA gene is now sequenced in a very wide range of organisms. Aligned sets of rRNA sequences are maintained at the Ribosomal Database Project (Cole *et al.* 2003) and the European Ribosomal RNA database (Wuyts *et al.* 2002). Phylogenies based on 16S rRNA have become a standard way of classifying bacteria (Woese, 1987; Olsen *et al.* 1994, Ludwig *et al.*, 1998). This gene is present in all organisms and it is rather slowly evolving, hence it can provide information at very deep levels of the evolutionary tree, such as the relationships between the major groups of bacteria. Nevertheless, it is unwise to base our understanding of evolution of whole organisms on just a single gene.

There is now a lot of evidence for horizontal transfer of genes between unrelated species of bacteria. An example of a well-studied set of genes is the aminoacyl-tRNA synthetases (Doolittle & Handy, 1998; Koonin *et al.* 2001) where specific anomalies occur in the phylogenetic trees that can best be interpreted as horizontal transfer events. Genes responsible for bacterial photosynthesis are also distributed between several separate groups, indicating probable horizontal transfer of these important functional genes (Raymond *et al.* 2003). Ochman *et al.* (2000) have estimated the amount of ‘foreign DNA’ arising from horizontal transfer in a range of bacterial genomes by looking for anomalies in statistical properties of the DNA sequences. The highest figures are 16.6% for *Synechocystis* and 12.8% for *E. coli* K12. A few percent of foreign DNA is detectable in most of the other genomes analysed. Even for the 16S gene itself, at least one case of horizontal transfer has been claimed (Yap *et al.* 1999).

If one accepts that horizontal transfer of all types of genes is common, one is led to the conclusion that the phylogenetic tree of bacteria is not a tree at all, but a tangle of interconnected branches (Doolittle, 2000). In the limit of very frequent horizontal transfer, every gene would have a different evolutionary tree. The 16S tree would be just one of these trees, of no more importance than any other. It might then be argued that there would be little point in doing phylogenetic studies with bacterial genes since no coherent picture would emerge. On the other hand, it has been argued that 16S may be one of a core set of genes that are very difficult to transfer. If we located such a core set of genes, we would expect that they should all conform to a consistent phylogenetic tree, and that this tree would really reflect the evolution of the organismal lineages. It has yet to be established whether such a core set exists, and which genes might be in it.

Woese (2000) defends the rRNA tree as being a valid representation of the organismal genealogy. He argues that horizontal transfer may have been common in the earliest organisms, but became progressively less important over evolutionary time, and progressively more confined to closely related species. Horizontal transfer within groups of closely related species would serve to define gene pools within which exchange of sequences is possible. However, it would not obscure the phylogenetic signal at deeper levels of the tree. According to this picture, the deeper levels of the bacterial tree should be consistent between genes, and the effort to elucidate the relationship between the major bacterial groups is worthwhile and potentially informative. Thus, the main issue is not whether a tree-like phylogeny exists, but whether genes exist that retain a reliable signal of phylogenetic relationships sufficiently far back in the past.

When we are dealing with genes from large multicellular organisms, the possibility of horizontal gene transfer can usually be discounted. Nevertheless, different genes sometimes give rise to different trees. This is indicative of the fact that phylogenetic methods are not perfect, and that sometimes the optimal trees according to the method used are not the biologically correct ones. In particular, this will be a problem if the model of evolution used in the likelihood calculations is too simple, or fails to account adequately for some important feature of the real sequences. Ludwig *et al.* (1998) compare phylogenies from several different genes and find many similarities but also important differences between the trees. They emphasize the limited phylogenetic information in any of these sequences, and interpret these results as simply illustrating our lack of certainty about the tree rather than as evidence for horizontal transfer. It is an unfortunate fact that phylogenetic results are sensitive to details of the methods used, the way the alignment is constructed and which set of species is included in the analysis. It is necessary to be sure we have overcome all these ‘normal’ problems of molecular phylogenetics before we can conclude that unusual events like horizontal transfer have occurred.

Since we now have many complete bacterial genomes, we have the potential to construct phylogenies from concatenated sets of genes. Suppose there is a moderate degree of horizontal transfer in the set of concatenated genes – say one event somewhere in the tree for each gene. Although the gene trees of each gene would then be different, this might actually do rather little to obscure the overall phylogenetic signal within the set of genes. One horizontal transfer corresponds to cutting one subtree from one of the gene trees and regrafting it somewhere else. This gene retains the signal that the species in the subtree form a clade, and retains information on the relationship between the species within the clade. It also retains the signal of all the other parts of the tree that were not moved. Therefore in a concatenated set of genes, if we assume that the horizontal transfer event was in a different place on the tree for each gene, then most genes will retain a reliable signal of the organismal tree in almost all parts of the tree. Reconstruction of the organismal tree from concatenated sets of genes therefore seems perfectly possible according to this picture.

At the outset of the study presented here, we will adopt the optimistic position that attempting to construct a tree for bacteria is a worthwhile exercise. We assume that there is a real phylogenetic signal to be found in bacterial evolution, either because there is a core set of genes, or because horizontal transfer is limited to closely related species, or because a substantial signal is still retained when different gene trees are regrafted in different places. In this paper we present phylogenetic trees using 16S and 23S rRNA and sets of tRNA genes from completely sequenced genomes. There have been several previous studies comparing 23S trees to 16S trees (De Rijk, 1995; Ludwig *et al.* 1995, 1998) although the number of available sequences for 23S is far fewer than for 16S. However, tRNA genes have not often been used for bacterial phylogenetics. One reason for this is that sequence information for complete sets of tRNAs from a wide range of species has only become available in the current era of complete genomes.

Individual tRNAs are very short, and a single tRNA gene tree is unlikely to give much resolution. However, concatenated sets of genes give a resulting alignment of comparable length to the rRNA genes. The genes for tRNAs are scattered throughout bacterial genomes. A small section taken randomly from a bacterial genome has a fairly large chance of containing a complete tRNA gene. Thus one might expect horizontal transfer of tRNAs to be relatively easy. On the other hand, tRNAs are essential genes found in all organisms, and are fairly slowly evolving, like rRNAs. Hence, it might be difficult to integrate a newly acquired tRNA gene into a genome. For these reasons, it is of interest to see how much phylogenetic signal is contained in tRNAs.

We have recently developed a set of phylogenetic programs known as PHASE (Jow et al. 2002) intended specifically to study the evolution of genes for RNA molecules with conserved secondary structure, such as tRNAs and rRNAs. Our programs use evolutionary models that account for the compensatory pairs of substitutions that occur in the paired regions of RNA secondary structures. We use Bayesian methods (Markov Chain Monte Carlo) for phylogenetic inference. We previously used these methods to study concatenated sets of rRNA and tRNA genes from mammalian mitochondrial genomes (Hudelot *et al.* 2003) and were able to obtain well-resolved trees showing the relationship between the orders of placental mammals. We also looked in detail at the evolution of tRNA-Leu genes in metazoan mitochondrial genomes (Higgs *et al.* 2003). There are two separate tRNAs for leucine, corresponding to two separate codon families in the genetic code. Our analysis showed at least five cases where one type of tRNA had evolved into the other due to a mutation in the anticodon. These cases showed up as sequences in anomalous positions in phylogenetic trees. Our ability to detect these anomalously positioned sequences in the mitochondrial tRNAs gives us hope that our methods may detect interesting anomalies in the bacterial tRNA trees that might arise due to horizontal transfer or due to anticodon mutations.

We have chosen the proteobacteria for this study. This is a fairly broad group, containing subdivisions that diverged quite early in evolution, as well as a diverse array of taxa at moderately deep levels (orders and families). We have a large number of complete genomes of proteobacteria (38 included here), thus enabling study of complete sets of tRNAs. The group also contains species of medical and agricultural relevance, and the establishment of a reliable phylogeny for these species is of interest in its own right. A good introduction to the classification of proteobacteria is given by De Ley (1992). The current classification is based principally on 16S rRNA phylogenies (Woese, 1987; Olsen *et al.* 1994; Ludwig *et al.* 1998) and recognizes five subdivisions of proteobacteria, labelled α, β, γ, δ and ε. These subdivisions are well defined clades, with the possible exception of the γ subdivision, which is sometimes found to be polyphyletic because the β group is contained within it. Gupta (2000) has also located insertions and deletions in protein coding genes that support several of the subdivisions and the relationships between them. The ‘backbone tree’ of bacteria from 16S rRNA available on the ribosomal database project web site (release 8.0 - Cole *et al.* 2003) gives the relationship between the subdivisions as (ε, (δ, (α, (β,γ)))), although this is not completely clear because some non-proteobacteria appear in a clade with the ε subdivision. The other possible arrangement, found for example by Olsen *et al.* (1994) is ((ε, δ), (α, (β,γ)))). Thus, the rooting of the proteobacteria is unclear, but there is general agreement on the existence of the clades β+γ and α+β+γ. We now proceed to the phylogenetic analysis of the RNA genes in proteobacteria, focussing initially on the species for which complete genomes are available.

## Methods

Table 1 lists the names of the species and the accession numbers for the sequences used in this study. These are classified according to the subdivision of the proteobacteria and the order. The classification follows the NCBI taxonomy available at http://www.ncbi.nlm.nih.gov/Taxonomy/taxonomyhome.html/. Most, but not all of these groupings are supported by the phylogenetic results in this paper.

Since our objective was to study complete sets of tRNA genes, we began with the set of 38 species for which complete genomes were available. Phylogenies were constructed for tRNAs, 16S and 23S rRNA genes for these species using the PHASE phylogenetic package (Jow *et al.* 2002) as described below. In order to assess the stability of the trees to the addition of extra species, a larger set of species was chosen by reference to the European Ribosomal RNA database (Wuyts *et al.* 2002). Species for which both 23S and 16S sequences were available were selected. We wished the selection to be as representative as possible of the full range of proteobacteria, but we wished to keep the numbers of species to a manageable level in order to avoid excessively long run times of the phylogenetic program. Therefore, a rapid distance matrix phylogeny was constructed for these sequences in order to reveal cases where there are large groups of extremely similar sequences which are of little phylogenetic interest. Some species were eliminated from each of these groups in order to prune down the tree. This resulted in a set of 96 species for which results are presented in this paper.

Separate multiple alignments were made for 16S, 23S and for each codon family of tRNA genes. ClustalX (Thompson *et al.* 1997) was used for an initial automatic alignment and this was then adjusted manually with the aid of GeneDoc (Nicholas & Nicholas, 1997) using inferred information about the secondary structure. To use the RNA-specific models of evolution in the PHASE package it is necessary to specify the conserved secondary structure in the alignment file. For the tRNAs, the clover leaf secondary structure can be added relatively easily by hand. For 16S and 23S rRNA, the secondary structures for the *E. coli* sequences (Cannone *et al.* 2002) were used as a reference. Pairs of sites that are indicated as forming a base pair in the *E. coli* structure were checked for conservation of this pair in the set to be studied. If a pair was found to be mismatching in more than 20% of sequences, this pair was removed from the structure and these two sites were treated as independently evolving single sites.

After aligning individual tRNA genes, we prepared a concatenated alignment of tRNAs that we will refer to as the ctRNA alignment. Table 2 lists the numbers of tRNA genes with each anticodon that occur in each genome. Genes present in some but not all species were excluded from the ctRNA alignment. The columns marked # in Table 2 have at least one gene in all the species, and these genes were included in the ctRNA alignment. Where more than one gene of a given anticodon is found in a species, one of these was selected at random. For the Arg-CGN codon family, there is no gene present in all species. However, the ACG gene is present in all but the ε species, and it was found that the GCG genes in the ε species are similar to the ACG genes in the other species. Therefore an alignment of ACG + GCG genes was included in the ctRNA alignment. Additionally, for the CAT anticodon (usually methionine) there are at least three genes in every species, and sequence analysis showed that these genes appear to be distinct, even though all have the same CAT anticodon. One of these is the initiator (fMet) tRNA, one is the normal Met tRNA and the third is actually an Ile tRNA rather than a Met. It is known from studies in *E. coli* that a modification of the C in the anticodon to a lysidine allows this tRNA to translate the AUA Ile codon (Muramatsu *et al.* 1988). This gene appears to be present in all the proteobacteria. We included all three CAT genes in the ctRNA alignment. Hence, though there are 27 marked columns in Table 2, there are 29 concatenated tRNAs for each species in the ctRNA alignment.

Before running the phylogenetic programs, sites in the alignment that contained large numbers of gaps were deleted. The numbers of single sites and paired sites retained in the alignments used are given in Table 3. Phylogenetic trees were produced using the Markov Chain Monte Carlo (MCMC) method implemented in PHASE and discussed in our previous papers (Jow et al. 2002; Hudelot et al. 2003; Higgs et al. 2003). The likelihood calculation used a general reversible four-state model for unpaired sites and a seven-state model for the paired sites. The possible states are AU, GU, GC, UA, UG, CG, and MM (a lumped state for all possible mismatches). The model used is referred to as 7D in the detailed comparison of models given by Savill *et al.* (2001), and is also fully described in the documentation to the PHASE package (http://www.bioinf.man.ac.uk/resources/phase). The rate parameters in both models are adjusted and optimized simultaneously during the MCMC simulation. Variability of rates between sites is allowed for in both models using a discrete approximation to a gamma distribution with four categories. A consensus tree was obtained from the set of trees generated in each MCMC run, ignoring the initial burn-in period. Repeat runs were done in order to check reproducibility of results and convergence of the MCMC sampling procedure. In order to get branch lengths on the consensus trees, the maximum likelihood tree was calculated for the consensus tree topology, keeping the topology fixed and allowing branch lengths and rate parameters to be optimised simultaneously.

**Comparison of phylogenies from 16S, 23S and concatenated tRNAs**

Figure 1 shows the consensus tree from the MCMC analysis for the ctRNA sequences. Only one of the four *E. coli* genomes and one of the 2 *Y. pestis* genomes has been included, since these are almost identical in the sequences of the tRNAs used. Similarly, *B. suis* is excluded since it is almost identical to *B. melitensis*. Figures 2 and 3 show the consensus trees for the 16S and 23S genes in the same species. We wish to ask how consistent these trees are with one another. Many of the clades in these trees are supported with 100% posterior probability, or with very high (*i.e.* $\geq$ 90%) support. For each sequence alignment we checked that several runs of the MCMC program beginning with different random trees converged to the same consensus tree. Hence any differences between the trees in the figures represent differences in the most likely trees supported by the different genes, and are not due to problems in equilibrating the MCMC simulations. This section highlights the similarities and differences between the trees, but we leave the interpretation of the differences to the next section.

The trees are presented as rooted with the ε subdivision as the earliest branching group. The root position cannot be confirmed from our work since the MCMC program deals with unrooted trees. The split between the $\varepsilon$ subdivision and the remaining $\alpha + \beta + \gamma$ subdivisions is supported at 100% in all three trees. There are no species in the $\delta$ subdivision in this set.

The clade comprising the $\alpha$ subdivision is also supported at 100% in all three trees. Within the $\alpha$ subdivision, the Rhizobiales and Rickettsiales clades are each supported at 100% in all three trees. However, the position of these groups with respect to the single *Caulobacter* sequence is different in each case. In the ctRNA tree, *Caulobacter* groups with the Rhizobiales with 100% support; in the 23S tree, *Caulobacter* groups with the Rickettsiales with 96% support; whilst in the 16S tree, *Caulobacter* branches first, but with weak support.

The clade comprising the $\beta + \gamma$ subdivisions is strongly supported**.** The clade for the $\beta$ subdivision is always supported at 100%, however, this sometimes this appears *within* the $\gamma$ subdivision. With the ctRNA set, the $\gamma$ subdivision is monophyletic, but with the other two genes, the $\gamma$ subdivision is polyphyletic due to the positioning of the $\beta$ subdivision (in different places for the two genes).

There are several well supported orders in the $\gamma$ subdivision: Xanthomonadales, Pseudomonadales, and Pasteurellales all have 100% support in all three trees and Vibrionales has 100% support in two of the three and □90% in the other one. In the ctRNA tree and 23S tree, the Enterobacteriales group has 100% support, but this order is polyphyletic in the 16S tree due to the positioning of Pasteurellales. A clade containing Enterobacteriales, Pasteurellales, Vibrionales and *Shewanella* exists in all three trees with 100% support. There is also support for a clade of Vibrionales + *Shewanella* in both ctRNA and 23S trees, but not 16S. A clade containing Xanthomonadales, Pseudomonadales and *Coxiella* exists in the trees for ctRNA (80% support) and 23S (99.9% support) but is not present in the 16S tree. Thus the relative positions of the groups within the $\gamma$ subdivision are not always well resolved, and are not always consistent between the three trees. In general there is closer agreement between the ctRNA and 23S trees than with 16S, and the 16S tree appears to be the least reliable of these three.

## Effect of adding further species

The previous section showed that there was broad agreement with most aspects of the phylogeny from the three types of gene but some disagreement on the details. The most important issues that arose were the position of *Caulobacter* within the α subdivision, the monophyly/polyphyly of the γ subdivision, and the monophyly/polyphyly of the Enterobacteriales. These problems are typical of those that arise in molecular phylogenetics. As pointed out in the introduction, even in cases where horizontal transfer can be ruled out, different genes and different methods sometimes give different trees. In the case of bacterial phylogenies, the possibility of horizontal transfer needs to be seriously considered, and it is necessary to distinguish whether an apparent inconsistency in a tree is due to a horizontal transfer or just a 'normal' phylogenetic problem. We attempted to do this by addition of extra species. It is known that trees can sometimes change when additional species are added. Increasing the numbers of species can serve to break up long internal branches and generally provides additional information that the phylogenetic program can use. We wish to know whether the discrepancies noted in the 38-species trees still remain in trees with the 96-species set. Clearly, any changes that arise when extra species are included have nothing to do with horizontal transfer.

Figures 4 and 5 show consensus trees for 16S and 23S genes with the 96-species set. We do not have complete sets of tRNAs for this larger set of species, hence no ctRNA tree is available. Trees have again been rooted with the ε division (but the root position cannot be concluded from this study). This time two representatives of the δ subdivision are included. In both trees 4 and 5, the ε and δ, subdivisions form monophyletic groups with 100% support; the α and β groups are also monophyletic with 100% support with the exception of the single species *Zoogloea ramigera*; but the γ subdivision is polyphyletic, as discussed further below. *Z. ramigera* is classified in the β subdivision according to the NCBI taxonomy (Table 1). According to the 16S tree (Fig 4) it is well within the β subdivision and is a sister group to *Ralstonia*. According to the 23S tree (Fig. 5) it is in the α subdivision in the middle of the Rhizobiales. These two positions are very far apart, and are separated by many well-supported nodes, therefore the result cannot be attributed to phylogenetic uncertainty. At first, we thought this was a clear candidate for horizontal transfer of a 23S gene into *Z. ramigera*. However, it has previously been noted (Shin *et al.* 1993) that the classification of the *Zoogloea* genus is questionable. Sequences of 16S rRNA are known from several strains denoted '*Z. ramigera*'. Some of these fall in the α subdivision and some in the β subdivision. We repeated the analysis using all available 16S sequences of *Z. ramigera*, and confirmed the observation of Shin *et al.* (1993). Hence the discrepancy between the positions of *Z. ramigera* in Figs. 4 and 5 is because these sequences are from different species, and not because of horizontal transfer.

In the larger data set, many additional species have been introduced to the α subdivision. Rhodobacterales, Rickettsiales and Caulobacterales are all well supported. Rhodospirillales is not consistently supported: *Rhodospirillum* is not with *Acetobacter* in either 16S or 23S trees and the position of *Rhodospirillum* is different in the two cases. The Rhizobiales form a well supported clade in the 23S tree, but are polyphyletic in the 16S tree because the Bradyrhizobiaceae are separated from the rest.

For the 16S gene, the relative positions of *Caulobacter*, Rhizobiales, and Rickettsiales is still the same in Fig. 4 as it was in Fig. 2, with *Caulobacter* branching earliest. For the 23S gene, *Caulobacter* is closer to the Rhizobiales in Fig. 5. This is different from Fig. 3, but the same as the ctRNA tree in Fig. 1. Addition of species has thus caused a rearrangement of the tree, indicating that we cannot be certain about the position of these groups. Most of the differences between the trees are due to nearest neighbour interchanges across short internal branches. Details like this are difficult to get right in many phylogenetic studies. Thus we cannot infer that the difference in the position of *Caulobacter* in the different trees is due to horizontal transfer.

The γ subdivision appeared to be polyphyletic in Figures 2 and 3. This is still true in Figures 4 and 5, although the position of the β clade within the γ subdivision is unstable. Comparing Figures 2 and 4 for the 16S gene shows that the relative positions of the Xanthomonadales, the β clade and *Coxiella* have changed due to the addition of species. Comparison of Figures 3 and 5 shows that the β clade has switched positions with Xanthomodales, Pseudomodadales and *Coxiella*. Since both 16S and 23S trees are unstable to the addition of species, we are forced to conclude that the position of the β subdivision cannot be relied on in either of these trees. In general we would expect that the trees with the larger number of species should be more reliable. There is a tendency for the β subdivision to move toward the base of the γ subdivision as more species are added. In Fig 5, the only species classified as γ that falls outside the β clade is *Acidithiobacillus ferrooxidans*. This species was not in the original set of complete genomes. Thus Fig. 5 is actually in agreement with Fig. 1 from the ctRNA alignment, in that the main groups within the γ subdivision form a clade that is completely separate from the β subdivision. A similar situation occurred in the results of Olsen *et al.* (1994) with 16S. Their selection of species contains *Xanthomonas, Coxiella* and *Pseudomonas*, and all these form a clade with the main body of the γ subdivision. In their tree, the γ subdivision is polyphyletic due to the position of *Ectothiorhodospira* and *Chromatium.* These species were unfortunately not in our set. Our tentative conclusion is, therefore, that if the γ subdivision is polyphyletic, it is only because of a few 'stragglers' like *Acidithiobacillus*, *Ectothiorhodospira* and *Chromatium*, and that there exists a 'true-γ' subdivision that contains all of the γ species that were in our original set of 38. This in turn implies that the ctRNA tree was more reliable than either 16S or 23S trees at the 38-species level.

On the question of the monophyly of the Enterobacteriales, it is only the 16S tree (Fig. 2) that casts doubt on this at the 38-species level. In Fig. 4, the Pasteurellales clade has moved well outside the Enterobacteriales, confirming that Enterobacteriales is monophyletic as expected. Confusingly, however, the additional species *Plesiomonas shigelloides*, classified within the Enterobacteriales, appears at the foot of that group in Fig. 4, but has wandered over to the foot of the Pasteurellales in Fig. 5. It is therefore not completely clear whether *Plesiomonas* merits inclusion within Enterobacteriales.

Also worthy of note is *Carsonella ruddii*, which is within Enterobacteriales in Fig. 5, but is in the *Zymobacter* group in Fig. 4. This is a very long jump in a single species, and must therefore be considered a serious candidate for a horizontal transfer event. Unfortunately, however, the issue is clouded by the very long branches leading to *C. ruddii* in both Figs. 4 and 5. In fact, *C. ruddii* is the most divergent species for both 16S and 23S genes. The position of solitary species on long branches is notoriously difficult to resolve, therefore we have rather little confidence in claiming that there has been a horizontal transfer in one of the *C. ruddii* genes.

The above discussion may appear somewhat negative, since we emphasized problems rather than successes with the phylogenetic results. We did this in order to underline the fact that the tree of proteobacteria is still far from being completely resolved, and hence that it is premature to argue for horizontal transfer as a general explanation of the discrepancies. However, there are many well-supported clades present in both Figs. 4 and 5, and in most cases the positions of these clades do not move very far from one tree to the next. We should not lose sight of the fact that phylogenetic analysis of these RNA genes actually tells us a lot. The results are typical of other phylogenetic problems where there are short internal branches at fairly deep levels of the tree that are very difficult to resolve fully.

**Phylogeny of Rhizobia**

There have been two recent papers dealing specifically with phylogeny within the Rhizobia (part of the α subdivision). Gaunt *et al.* (2001) compared the tree derived from 16S rRNA with those from atpD and recA genes. They concluded that there is a broad phylogenetic agreement between these trees, and that these genes belong to a set of housekeeping genes for which there is a robust and consistent phylogeny. On the other hand, van Berkum *et al.* (2003) compared 16S and 23S phylogenies using a similar set of species, and concluded that they were significantly different from one another. Part of this apparent difference in conclusions can be attributed to the level at which these authors base their discussions. Whereas van Berkum *et al.* emphasize the possibility of recombination, gene conversion and horizontal transfer within genera, Gaunt *et al.* emphasize the consistency of the phylogeny above the genes level, but acknowledge that there is some evidence for recombination within genera. Thus, both groups suggest that phylogenetic inconsistencies are more likely to arise at a narrow phylogenetic level.

Our results up to this point apply to a broad level – the whole of the proteobacteria. We therefore wished to look at a more closely related group of species to see if the same conclusions apply. We chose the same set of species from the Rhizobia as van Berkum *et al.* (2003), and we also used the same species as outgroup (*Rhodobacter sphaeroides*), although the position of *R. sphaeroides* with respect to the Rhizobia is not consistent in our broad-scale results (Figures 4 and 5). Neither van Berkum *et al.* (2003) or Gaunt *et al.* (2001) included *Brucella* or *Bartonella* in their study. However, our results for both 16S and 23S agree in placing these species closer to the *Agrobacterium / Rhizobium* group that they are to the *Bradyrhizobium / Rhodopseudomonas* group; therefore we included them in the analysis.

Our results for the Rhizobia using 16S and 23S genes are presented in Figures 6 and 7, respectively. These may be compared with van Berkum *et al.* (2003) Figures 1A and 1B and with Gaunt *et al.* (2001) Figure 1. In both genes, the primary split is between the clade labelled 1 and the rest. Clade 1 contains *A. caulinodans* as an early branching species, together with a cluster involving *Bradyrhizobium*, *Afipia*, and *Rhodopseudomonas*. This is consistent with both previous studies. The branching order within this cluster is not equivalent in Figures 6 and 7, as there are several nodes with 90-100% posterior probability that differ between the trees. Nevertheless, the internal branches within this cluster are all rather small. We interpret this as suggestive of some degree of non-treelike evolution among this group, but consider the evidence to be very weak. (Note that *B. denitrificans* was omitted from the 23S tree because this sequence is incomplete.)

Clade 2 is present in both trees. The branching order of the four species is different, but is not strongly supported in Fig. 7, in any case. Clade 3 is present in Fig. 6, but is polyphyletic in Fig. 7. However, the relevant internal branches in Fig. 7 are extremely short and are not strongly supported. Thus there is no strong evidence of inconsistency here. The combined clade 2+3 is present in both trees, and this supports the proposal for merging *Agrobacterium* and *Rhizobium* into one genus (see discussion in Gaunt *et al.*, 2001). Again, these results are very similar to those of van Berkum *et al.* (2003).

Clades 4 and 5 correspond to the genera *Sinorhizobium* and *Mesorhizobium*. These clades are consistent between the two figures, and also occur in previous studies. In Fig. 7, there is a large combined clade 2+3+4 with 100% support, whereas in Fig. 6, clade 4 appears closer to clade 5. The nodes defining the relationships between clades 2-5 do not have strong support in Fig 6, however, so at best this is very weak evidence for inconsistency of these gene trees.

In addition to these well-defined clades, there are several species that move a substantial distance between Figs. 6 and 7. In Fig. 7, *Phyllobacterium, Ochrobactrum, Bartonella* and *Brucella* form clade 6, whereas these species are definitely polyphyletic in Fig. 6. The most important inconsistency here is the position of *Bartonella*: there is 100% support for a pairing with *Brucella* in Fig. 7, and a 90% support for a pairing with *Phyllobacterium* and *Mesorhizobium* in Fig. 6. In order to create clade 6 from the tree in Fig.6 we need to shift the *Brucella / Ochrobactrum* group towards the top of the tree. This involves only a small number of nearest neighbour interchanges across nodes that are not strongly supported. The evidence of inconsistency between trees is moderately strong in this respect, but we are still left with a doubt that species like *Bartonella* and *Brucella* may be 'problem' species that are difficult to position due to factors unrelated to horizontal transfer.

In general, the deeper levels of the tree are better supported with the 23S gene than with 16S. We therefore consider the clades 6, (5+6), and (2+3+4) as proposals that should be considered seriously by those interested in relationships among the Rhizobia.

**Analysis of individual tRNA trees**

We also carried out separate phylogenetic analyses of the tRNAs for each amino acid. Determination of the rate parameters for both the four-state and seven-state models simultaneously from single tRNA sequences was found to be unreliable because the sequences are short. For this reason, we obtained the maximum likelihood rate parameters for both models from the ctRNA alignment, and the parameters were fixed at these values during the MCMC runs for the single genes. We also analysed each gene using only the four-state model and ignoring the secondary structure.

We will discuss the case of arginine tRNAs, since this raises several important issues. All species possess genes for both the CGN and AGR codon family (see Table 2). A mutation in the anticodon could cause a gene to switch from one family to the other. A first question is therefore to ask if these genes are evolutionarily distinct from one another, or if switches between families can be observed. The pattern of presence and absence of genes among the arginine tRNAs is also interesting: TCT is present in all species; ACG and CCG are present in all (or almost all) except the ε subdivision; GCG is present only in the ε subdivision; TCG is present only in the ε subdivision and *E. coli*; and CCT has a very sporadic pattern of presence and absence. The general question posed by these data is to ascertain how the genes came to be present in certain genomes but not others.

Our phylogenies with single tRNA-Arg sequences are not resolved at the species level, and are therefore not shown. However, they do allow a few key points to be concluded with reasonable certainty. The genes from the two different codon families fall into two separate groups that are clearly distinguishable from one another, with the exception of the TCG genes from *E. coli*. These genes fall within a well-defined clade of genes with TCT and CCT anticodons, and appear to be most similar to TCT sequences from *E. coli* and *S. typhimurum*.

This suggests that the TCG genes in *E. coli* have arisen as a result of an anticodon mutation in a TCT gene of the same species. This causes a gene to switch between codon families.

In contrast, the TCG and GCG genes in the ε subdivision appear to have arisen as a result of anticodon mutations *within* a codon family. Within this family, the ACG and CCG genes are fairly well separated from one another, indicating that these genes have probably been distinct genes throughout the evolution of the proteobacteria. Both the TCG and GCG genes of the ε subdivision appear to be close to ACG genes rather than CCG, and we therefore conclude that they arose via anticodon mutations in ACG genes. The ACG gene is absent in the ε subdivision, which could be due to a deletion or due to the same mutation event that created the TCG and/or GCG gene. Similarly, the CCG gene is absent in the ε subdivision, which presumably is due to a gene deletion. Suspiciously, CCG is also absent in *C. burnetii* and one of the *N. meningitidis* strains. This could be due to two further independent gene deletions, but might indicate a failure to locate this gene on the genome (we did not carry out independent whole-genome searches for tRNAs).

The CCT gene has a sporadic pattern of presence and absence, and clearly this is not a problem of failure to locate the gene in all of these spsecies. Our interpretation is that there have been multiple independent deletions of this gene. If the ctRNA tree is a true representation of the organismal phylogeny, a total of 5 independent deletions of the same gene would have to have occurred. It might at first seem non-parsimonious to argue for 5 deletions in one gene whilst other genes have not been deleted at all. However, deletions are not random. We know that some organisms manage perfectly well with only TCT genes, because the UCU tRNAs transcribed from these genes are able to translate both AGG and AGA codons. The possession of a CCT gene might therefore been seen as a 'luxury' that might slightly improve translation efficiency of the AGG codon, but would be non-essential. In contrast, the TCT gene *is* essential, and hence no deletions of TCT genes are observed in surviving organisms. There are several other examples of genes with sporadic presence/absence patterns in table 2 (*e.g.* CTG, CAA, GGG *etc.*) where multiple deletions seem to have occurred. This is the sort of pattern we would expect to find for genes that are non-essential, particularly since there are several species in this set with substantially reduced genome size (*e.g. Buchnera* and *Rickettsia*) where we know many gene deletions have occurred.

We analysed each of the tRNA gene families in a similar way to the arginine tRNAs described above. We paid special attention to leucine and serine tRNAs, since these are the other two amino acids for which there are two separate codon families in the standard genetic code. We tentatively conclude that genes from the two families have remained separate for both these amino acids and that there has been no switching between codon families. This is in contrast to tRNA-Leu genes in animal mitochondria, where this occurs several times (Higgs *et al.* 2003). In carrying out the individual tRNA phylogenies, we hoped to be able to determine a most likely scenario for the presence/absence patterns of each gene. Unfortunately, it is possible to explain any observed pattern with many different scenarios involving gene deletion, origination of new genes, and horizontal gene transfer. Coupled with this, the resolution of the individual tRNA trees was quite poor in many cases. Although our results are suggestive, the conclusions we are able to draw are not sufficiently firm to justify a detailed presentation of results for every amino acid.

**Discussion**

The way one interprets the presence/absence pattern for a gene depends on one's prior assumptions about the likelihood of gene deletions, duplications, new gene origins and horizontal transfer. It appears that deletion of non-essential tRNA genes is frequent in this data set. This is suggested by the sporadic pattern of absences in many of the columns of Table 2, as discussed above. In addition to this, the number of copies of the genes that are present in all species varies considerably between species, which suggests that both gene duplication and deletion are frequent. For protein coding genes, the origination of a new gene seems an improbable event, and one would be very unlikely to propose a scenario with two independent origins of the same gene. However, for tRNAs, a gene with an anticodon not previously present can arise easily by a point mutation of another tRNA that is present in multiple copies. This means that multiple independent origins of some of the tRNA genes is also perfectly possible as an interpretation of the presence/absence patterns in Table 2 (*e.g.* in the previous section, we proposed two independence origins for the TCG gene). In addition to anticodon mutations between tRNAs for the same amino acid (such as those in tRNA-Arg genes discussed above), Saks *et al.* (1998) have argued that anticodon mutations can occur between genes for different amino acids. We would agree that this seems perfectly possible, but the lack of resolution of the individual tRNA trees makes it difficult to point to particular cases where this has occurred.

Given that gene duplication, deletion and origination via anticodon mutation all seem to occur frequently in tRNAs, what can we say about the rate of horizontal transfer? The general agreement of the ctRNA tree with the 16S and 23S rRNA trees suggests to us that horizontal transfer is rarer than any of these other events. Our results show that the subdivisions ε, δ, α, and (β+γ) are well defined, and we can be reasonably confident in ruling out horizontal transfer between subdivisions. The question of monophyly/paraphyly of the γ subdivision is not resolved, but in our view, this problem is not related to horizontal transfer. Furthermore, within the subdivisions, there are several well-defined clades at the level of orders and families. At the detailed level (within a genus or between closely related genera, such as among the Rhizobia), we found several cases where a moderate degree of inconsistency arises between gene trees. This is suggestive of non-treelike evolution, and we cannot rule out horizontal gene transfer as an explanation. However, we did not find cases where we are completely confident in asserting that a horizontal transfer has occurred. For those of us who are not specialists on any one genus, it is the large scale phylogenetic relationships that are most interesting. If horizontal transfer *is* occurring, then it is doing so in a way that makes remarkably little difference to the most interesting aspects of bacterial phylogeny.

We will conclude by emphasizing that tRNA sequences appear to be very reliable indicators of bacterial phylogeny. The ctRNA tree is at least as well resolved as the 16S and 23S trees. Although some degree of effort is required to handle large numbers of small sequences, the conservation of the tRNA secondary structure means that it is possible to get very reliable alignments of tRNA genes. In our view the ctRNA alignment is extremely informative. Both the 16S and 23S trees changed somewhat when extra species were added, and became somewhat closer to the ctRNA tree. If making a prediction of the ‘correct’ phylogeny for the 38 species with complete genomes, we would attach higher confidence to the ctRNA tree (Fig. 1) than either of the rRNA trees (Figs. 2 and 3).

**Acknowledgements**

This work was supported by the Canada Research Chairs programme. We thank Vivek Gowri-Shankar for advice regarding the PHASE software.

## References

Cannone J.J., Subramanian S., Schnare M.N., Collett J.R., D'Souza L.M., Du Y., Feng B., Lin N., Madabusi L.V., Muller K.M., Pande N., Shang Z., Yu N., and Gutell R.R. (2002). The Comparative RNA Web (CRW) Site: An Online Database of Comparative Sequence and Structure Information for Ribosomal, Intron, and other RNAs. *BioMed Central Bioinformatics*, **3**:2. http://www.rna.icmb.utexas.edu/

Cole, J.R., Chai, B., Marsh, T.L., Farris, R.J., Wang, Q., Kulam, S.A., Chandra, S., McGarrell, D.M., Schmidt, T.M., Garrity, G.M. & Tiedje, J.M. (2003) The ribosomal database project (RDP-II): previewing a new autoaligner that allows regular updates and the new prokaryotic taxonomy. http://rdp.cme.msu.edu/html/

De Ley, J (1992) The proteobacteria: ribosomal RNA cistron similarities and bacterial taxonomy. In: The Prokaryotes (Balows, A., Trüper, H.G., Dworkin, M., Harder, W. and Schleifer, K.H., Eds.), pp. 2111–2140. Springer-Verlag, New York.

De Rijk, P., Van de Peer, Y., Van den Broeck, I. (1995) Evolution according to large ribosomal subunit RNA. *J. Mol. Evol.* **41**, 366-375.

Doolittle, R.F. & Handy, J. (1998) Evolutionary anomalies among the aminoacyl-tRNA synthetases. *Curr. Op. Genet. Dev.* **8**, 630-636.

Doolittle, W.F. (2000) Uprooting the tree of life. *Sci. Am.* **282**, 90-95.

Gaunt, M.W., Turner, S.L., Tigottier-Gois, L., Lloyd-Macgilp, S.A. & Young, J.P.W. (2001) Phylogenies of *atpD* and *recA* support the small subunit rRNA-based classification of rhizobia. *Int. J. Syst. Evol. Microbiol.* **51**, 2037-2048.

Gupta, R.S. (2000) The phylogeny of proteobacteria: relationships to other eubacterial phyla and eukaryotes. *FEMS Microbiol Rev* **24**, 367-402.

Higgs, P.G., Jameson, D., Jow, H. & Rattray, M. (2003) The evolution of tRNA-Leucine genes in animal mitochondrial genomes. *J. Mol. Evol.* **57**, 435-445.

Hudelot, C., Gowri-Shankar, V., Jow, H., Rattray, M. & Higgs, P.G. (2003) RNA-based Phylogenetic Methods: Application to Mammalian Mitochondrial RNA Sequences. *Mol. Phyl. Evol.* **28**, 241-252

Jow H, Hudelot C, Rattray M, Higgs PG (2002) Bayesian phylogenetics using an RNA substitution model applied to early mammalian evolution. Mol. Biol. Evol. 19:1591-1601. Source code and documentation available at http://www.bioinf.man.ac.uk/resources/phase .

Koonin, E.V., Makarova, K.S. & Aravind, L. (2001) Horizontal Gene Transfer in Prokaryotes: Quantification and Classification. *Annu. Rev. Microbiol.* **55**, 709-742.

Ludwig, W., Rosselo-Mora, R., Aznar, R., Klugbauer, S., Spring, S., Reetz, K., Beimfohr, C., Brockmann, E., Kirchhof, G., Dom, S., Bachleitner, N., Klugbauer, N., Springer, N., Lane, D., Nietupsky, R., Weizenegger, M., Schleifer, K.H. (1995) Comparative sequence analysis of 23S rubosomal RNA from proteobacteria. *Syst. Appl. Microbiol.* **18**, 164-188.

Ludwig, W., Strunk, O., Klugbauer, S., Klugbauer, N., Weizenegger, M., Neumaier, J., Bachleitner, M., Schleifer, K.H. (1998) Bacterial phylogeny based on comparative sequence analysis. *Electrophoresis* **19**, 554-568.

Muramatsu, T., Yokoyama, S., Horie, N., Matsuda, A., Ueda, T., Yamaizumi, Z., Kuchino, Y., Nishimura, S., & Miyazawa, T. (1988) A Novel Lysine-substituted Nucleoside in the First Position of the Anticodon of Minor Isoleucine tRNA from *Escherichia coli*. *J. Biol. Chem.* **263**, 9261-9267.

Nicholas KB, Nicholas HB Jr. (1997) Genedoc: a tool for editing and annotating multiple sequences alignments, http://www.psc.edu/biomed/genedoc

Ochman, H., Lawrence, J.G. R Groisman, E.A. (2000) Lateral gene transfer and the nature of bacterial innovation. *Nature* **405**, 299-304.

Olsen, G. J., Woese, C. R., and Overbeek, R. 1994. The winds of (evolutionary) change: Breathing new life into microbiology, J. Bacteriol. 176, 1–6.

Raymond J, Zhaxybayeva O, Gogarten JP, Blankenship RE (2003) Evolution of photosynthetic prokaryotes: a maximum-likelihood mapping approach. *Phil. Trans. Roy. Soc.* **B 358**, 223-230.

Saks ME, Sampson JR and Abelson J (1998) Evolution of a Transfer RNA Gene Through a Point Mutation in the Anticodon Science, 279, 1665-1670

Savill, N.J., Hoyle, D.C. & Higgs, P.G. (2001) RNA sequence evolution with secondary structure constraints: Comparison of substitution rate models using maximum likelihood methods. *Genetics* **157**, 399-411.

Shin, Y.K., Hirashi, A. & Sugiyama, J. (1993) Molecular systematics of the genus Zoogloea and emendation of the genus. *Int. J. Syst. Bacteriol.* **43**, 826-831.

Thompson JD, Gibson TJ, Plewniak F, Jeanmougin F, Higgins DG (1997) The ClustalX windows interface: flexible strategies for multiple sequence alignment aided by quality analysis tools. Nucl. Acids Res. 24:4876-4882.

van Berkum, P., Terefework, Z., Paulin, L., Suomalainen, S., Lindstrom, K., and Eardly, B.D. (2003) Discordant Phylogenies within the *rrn* Loci of Rhizobia. *J. Bacteriol.* **185**, 2988-2998.

Woese CR (1987) Bacterial evolution. *Microbiol. Rev.* 51:221–271.

Wuyts, J., Van de Peer, Y., Winkelmans, T., De Wachter R. (2002) The European database on small subunit ribosomal RNA. *Nucleic Acids Res.* **30,** 183-185. http://oberon.fvms.ugent.be:8080/rRNA/ssu/index.html

Yap, W.H., Zhang, Z. & Wang, Y. (1999) Distinct types of rRNA operons exist in the genome of the actinomycete *Thermomonospora chromogena* and evidence for the horizontal transfer of an entire rRNA operon. *J. Bacteriol.* **181**, 5201-5209.

## Figure Captions

Figure 1 – Consensus phylogeny of the 29 concatenated tRNAs for the 38 species set. Branch lengths are the maximum likelihood values for the consensus tree topology. Nodes supported with 100% posterior probability in the Bayesian analysis are marked ● and those with posterior probability □90% are marked ○. The same notation is used in figures 2 – 7.

Figure 2 – Consensus phylogeny of 16S rRNA for the 38 species set.

Figure 3 – Consensus phylogeny of 23S rRNA for the 38 species set.

Figure 4 – Consensus phylogeny of 16S rRNA for the 96 species set.

Figure 5 – Consensus phylogeny of 23S rRNA for the 96 species set.

Figure 6 – Consensus phylogeny of 16S rRNA for the Rhizobiales.

Figure 7 – Consensus phylogeny of 23S rRNA for the Rhizobiales.

Table 1. Classification of the proteobacteria included in this study according to the NCBI taxomony site http://www.ncbi.nlm.nih.gov/Taxonomy/taxonomyhome.html.
The symbol ● indicates that the species is included in the set of 38 complete genomes (Figures1, 2 and 3), * indicates that the species is included in the narrow-level study of the Rhizobia (Figures 6 and 7). The initials p. e. & s. e. stand for primary endosymbiont & secondary endosymbiont.

| Subdivision / Order | Species | Accession number |
|---|---|---|
| **Epsilon** | | |
| Campylobacterales | Campylobacter coli | U09611, M59073 |
| | Campylobacter jejuni ● | NC_002163 |
| | Helicobacter pylori J99 ● | NC_000921 |
| | Helicobacter pylori 26695 ● | NC_000915 |
| | Wolinella succinogenes | NC_005090 |
| **Delta** | | |
| Myxococcales | Nannocystis exedens | X87286, M94279 |
| | Stigmatella aurantiaca | X87291, M94281 |
| **Alpha** | | |
| Rhizobiales | Afipia felis * | AF003937, AF338177 |
| | Agrobacterium radiobacter | AF209074, AJ130719 |
| | Agrobacterium rhizogenes * | D14501, AF208480 |
| | Agrobacterium rubi * | AF208483, D14503 |
| | Agrobacterium tumefaciens ● * | NC_003062 |
| | Agrobacterium vitis * | AF209071, AJ389912 |
| | Azorhizobium caulinodans * | X94200, AY244367 |
| | Bartonella bacilliformis * | L39095, X60042 |
| | Blastobacter denitrificans * | S46917 |
| | Bradyrhizobium elkanii * | AF237422, U35000 |
| | Bradyrhizobium japonicum * | X66024, Z35330 |
| | Bradyrhizobium japonicum USDA 110 * | NC_004463 |
| | Bradyrhizobium lupini | X87283, X87273 |
| | Brucella melitensis ● * | NC_003317 |
| | Brucella suis ● * | NC_004311 |
| | Mesorhizobium amorphae * | AF041442, AY244358 |
| | Mesorhizobium ciceri * | U07934, Z79618 |
| | Mesorhizobium huakuii * | D13431, AY244366 |
| | Mesorhizobium loti * | X67229, AY244364 |
| | Mycoplana dimorpha * | D12786, D12786 |
| | Ochrobactrum anthropi * | D12794, AY244379 |
| | Phyllobacterium myrsinacearum * | D12789, AY244380 |
| | Rhizobium etli * | U28916, AY244372 |
| | Rhizobium galegae * | AF207783, AY167831 |
| | Rhizobium gallicum * | U86343, AY244362 |
| | Rhizobium huautlense * | AF025852, AY244375 |
| | Rhizobium leguminosarum * | AF207782, X91211 |
| | Rhizobium tropici * | AF208479, X77125 |
| | Rhodopseudomonas palustris * | AF184625 |
| | Sinorhizobium fredii * | X67231, AY244360 |
| | Sinorhizobium kostiense * | Z78203, AY244382 |
| | Sinorhizobium meliloti • * | NC_003047 |
| | Sinorhizobium saheli * | X68390, AY244368 |
| | Sinorhizobium arboris * | Z78204, AY244381 |
| | Sinorhizobium terangae * | X68387, AY244369 |
| Rickettsiales | Rickettsia canadensis | AJ133712, L36104 |
| | Rickettsia conorii • | NC_003103 |
| | Rickettsia prowazekii • | NC_000963 |
| | Rickettsia rhipicephali | Y13128, L36216 |
| | Wolbachia pipientis | U23710, AJ306315 |
| Caulobacterales | Brevundimonas diminuta | X87288, X87274 |
| | Caulobacter crescentus • | NC_002696 |
| Rhodospirillales | Acetobacter europaeus | X89771, Z21936 |
| | Acetobacter intermedius | Y14680, Y14694 |
| | Acetobacter xylinum | X89812, X75619 |
| | Rhodospirillum rubrum | X87290, X87278 |
| Rhodobacterales | Paracoccus denitrificans | X87287, X69159 |
| | Rhodobacter capsulatus | X06485, D16427 |
| | Rhodobacter sphaeroides * | X53853 |
| Sphingomonadales | Zymomonas mobilis | AF086792 |
| **Beta** | | |
| Burkholderiales | Alcaligenes faecalis | X87282, AJ550276 |
| | Burkholderia gladioli | Y17182, L28156 |
| | Burkholderia mallei | Y17183, L28158 |
| | Bordetella avium | X70370, AF177666 |
| | Bordetella bronchiseptica | X70371, X57026 |
| | Bordetella parapertussis | X68368, AJ278450 |
| | Ralstonia pickettii | AF012421,AJ270260 |
| | Ralstonia solanacearum • | NC_003295 |
| Neisseriales | Neisseria gonorrhoeae | X67293, AJ247239 |
| | Neisseria meningitidis MC58 • | NC_003112 |
| | Neisseria meningitidis Z2491 • | NC_003116 |
| Rhodocyclales | Zoogloea ramigera | X88894, X74914 |
| **Gamma** | | |
| Acidithiobacillales | Acidithiobacillus ferrooxidans | U18089 |
| Legionellales | Coxiella burnetii • | X79704, D89799 |
| Thiotrichales | Leucothrix mucor | X87285, X87277 |
| Xanthomonadales | Xanthonomas axonopodis • | NC_003919 |
| | Xanthonomas campestris • | NC_003902 |
| | Xylella fastidiosa • | NC_002488 |
| | Xylella fastidiosa Temecula1 • | NC_004556 |
| Oceanospirillales | Carsonella ruddii | AF211123 |
| | p. e. of Bemisia argentifolii | AF211870 |
| | Zymobacter palmae | AF211871 |
| Pseudomonadales | Acinetobacter calcoaceticus | X87280, AJ247199 |
| | Pseudomonas aeruginosa • | NC_002516 |
| | Pseudomonas fluorescens | AF134704 |
| | Pseudomonas stutzeri | U65012 |
| | Pseudomonas syringae • | NC_004578 |
| Aeromonadales | Aeromonas hydrophila | X67946, X87271 |
| | Ruminobacter amylophilus | X06765, AB004908 |
| Alteromonadales | Shewanella oneidensis • | NC_004347 |
| Vibrionales | Vibrio cholerae • | NC_002505 |
| | Vibrio parahaemolyticus • | NC_004603 |
| | Vibrio vulnificus • | X74727, X87293 |
| Pasteurellales | Haemophilus influenzae Rd • | NC_000907 |
| | Haemophilus influenzae Rd A | U32742 |
| | Pasteurella multicoda • | NC_002663 |
| Enterobacteriales | Buchnera aphidicola (S.g.) • | NC_004061 |
| | Buchnera aphidicola (B.p.) • | NC_004545 |
| | Buchnera aphidicola (A.p.) • | NC_002528 |
| | Citrobacter freundii | U77928, AJ514240 |
| | Escherichia coli CFT073 • | NC_004431 |
| | Escherichia coli K-12 • | NC_000913 |
| | E. coli O157:H7 EDL933 • | NC_002655 |
| | E. coli O157:H7 • | NC_002695 |
| | Klebsiella pneumoniae | X87284, X87276 |
| | Plesiomonas shigelloides | X65487, X74688 |
| | s. e. of Aphalaroida inermis | AF263556 |
| | s. e. of Bactericera cockerelli | AF263557 |
| | s. e. of Blastopsylla occidentalis | AF263558 |
| | s. e. of Cacopsylla myrthi | AF263559 |
| | s. e. of Calophya schini | AF263560 |
| | s. e. of Glycaspis brimblecombei | AF263561 |
| | s. e. of Heteropsylla texana | AF263562 |
| | Salmonella bongori | U77927,AF029227 |
| | Salmonella enterica • | NC_003198 |
| | Salmonella typhimurium • | NC_003197 |
| | Shigella flexneri • | NC_004337 |
| | Yersinia enterocolitica | U77925, X68674 |
| | Yersinia pestis C092 • | NC_003143 |
| | Yersinia pestis KIM • | NC_004088 |

Table 2. Numbers of tRNAs in each genome for each anticodon

| | Ala | | | Arg-CGN | | | | Arg-AGR | | Asn | Asp | Cys | Gln | | Glu | | Gly | | | His |
|---|---|---|---|---|---|---|---|---|---|---|---|---|---|---|---|---|---|---|---|---|
| | GGC | TGC | CGC | ACG | GCG | TCG | CCG | TCT | CCT | GTT | GTC | GCA | TTG | CTG | TTC | CTC | GCC | TCC | CCC | GTG |
| C. jejuni | 1 | 3 | 0 | 0 | 1 | 1 | 0 | 1 | 1 | 1 | 2 | 1 | 1 | 0 | 1 | 0 | 2 | 1 | 0 | 1 |
| H. pylori 26695 | 1 | 1 | 0 | 0 | 1 | 1 | 0 | 1 | 1 | 1 | 1 | 1 | 1 | 0 | 2 | 0 | 1 | 1 | 0 | 1 |
| H. pylori J99 | 1 | 1 | 0 | 0 | 1 | 1 | 0 | 1 | 1 | 1 | 1 | 1 | 1 | 0 | 2 | 0 | 1 | 1 | 0 | 1 |
| | | | | | | | | | | | | | | | | | | | | |
| A. tumefaciens | 1 | 4 | 0 | 1 | 0 | 0 | 1 | 1 | 0 | 1 | 2 | 1 | 1 | 1 | 2 | 0 | 2 | 2 | 0 | 1 |
| S. meliloti | 1 | 3 | 1 | 1 | 0 | 0 | 1 | 1 | 1 | 1 | 2 | 1 | 1 | 1 | 2 | 1 | 1 | 1 | 1 | 1 |
| B. melitensis | 1 | 3 | 1 | 1 | 0 | 0 | 1 | 1 | 1 | 1 | 2 | 1 | 1 | 1 | 1 | 1 | 2 | 1 | 1 | 1 |
| B. sui | 1 | 3 | 1 | 1 | 0 | 0 | 1 | 1 | 1 | 1 | 2 | 1 | 1 | 1 | 2 | 1 | 2 | 1 | 1 | 1 |
| R. conorii | 1 | 1 | 0 | 1 | 0 | 0 | 1 | 1 | 0 | 1 | 1 | 1 | 1 | 0 | 1 | 0 | 1 | 1 | 0 | 1 |
| R. prowazekii | 1 | 1 | 0 | 1 | 0 | 0 | 1 | 1 | 0 | 1 | 1 | 1 | 1 | 0 | 1 | 0 | 1 | 1 | 0 | 1 |
| C. crescentus | 1 | 2 | 1 | 1 | 0 | 0 | 1 | 1 | 1 | 2 | 2 | 1 | 1 | 0 | 2 | 0 | 2 | 1 | 1 | 1 |
| | | | | | | | | | | | | | | | | | | | | |
| R. solanacearum | 2 | 4 | 1 | 1 | 0 | 0 | 1 | 1 | 1 | 1 | 3 | 2 | 1 | 0 | 2 | 0 | 2 | 1 | 1 | 1 |
| N. meningitidis MC58 | 1 | 4 | 0 | 2 | 0 | 0 | 1 | 1 | 1 | 2 | 2 | 1 | 3 | 0 | 2 | 0 | 4 | 1 | 0 | 1 |
| N. meningitidis Z2491 | 1 | 3 | 0 | 1 | 0 | 0 | 0 | 1 | 1 | 0 | 2 | 1 | 3 | 0 | 1 | 0 | 2 | 1 | 0 | 1 |
| | | | | | | | | | | | | | | | | | | | | |
| C. burnetii | 1 | 1 | 0 | 2 | 0 | 0 | 0 | 1 | 1 | 1 | 1 | 1 | 1 | 0 | 1 | 1 | 1 | 1 | 1 | 1 |
| X. fastidiosa | 1 | 2 | 1 | 1 | 0 | 0 | 1 | 1 | 1 | 1 | 1 | 1 | 1 | 1 | 1 | 1 | 2 | 1 | 1 | 1 |
| X. fastidiosa T. | 1 | 2 | 1 | 1 | 0 | 0 | 1 | 1 | 1 | 1 | 1 | 1 | 1 | 1 | 1 | 1 | 2 | 1 | 1 | 1 |
| X. axonopodis | 2 | 3 | 1 | 2 | 0 | 0 | 1 | 1 | 1 | 1 | 2 | 1 | 1 | 1 | 2 | 1 | 2 | 1 | 1 | 1 |
| X. campestris | 2 | 2 | 1 | 2 | 0 | 0 | 1 | 1 | 1 | 1 | 2 | 1 | 1 | 1 | 2 | 1 | 2 | 1 | 1 | 1 |
| P. aeruginosa | 2 | 4 | 0 | 3 | 0 | 0 | 1 | 1 | 1 | 2 | 4 | 1 | 1 | 0 | 3 | 0 | 3 | 1 | 1 | 2 |
| P. syringae | 2 | 5 | 0 | 2 | 0 | 0 | 1 | 1 | 1 | 2 | 3 | 1 | 1 | 0 | 4 | 0 | 3 | 1 | 1 | 1 |
| S. oneidensis | 2 | 3 | 0 | 6 | 0 | 0 | 1 | 1 | 1 | 5 | 4 | 2 | 3 | 0 | 6 | 0 | 6 | 1 | 0 | 2 |
| V. cholerae | 1 | 4 | 0 | 6 | 0 | 0 | 1 | 1 | 0 | 4 | 5 | 3 | 5 | 0 | 4 | 0 | 6 | 2 | 0 | 2 |
| V. parahaemolyticus | 1 | 4 | 0 | 8 | 0 | 0 | 1 | 1 | 1 | 5 | 6 | 4 | 6 | 0 | 6 | 0 | 11 | 2 | 0 | 2 |
| V. vulnificus | 1 | 5 | 0 | 6 | 0 | 0 | 1 | 1 | 1 | 5 | 6 | 3 | 4 | 0 | 5 | 0 | 7 | 2 | 0 | 2 |
| H. influenzae Rd | 1 | 3 | 0 | 2 | 0 | 0 | 1 | 1 | 0 | 2 | 3 | 1 | 2 | 0 | 3 | 0 | 3 | 1 | 0 | 1 |
| P. multicoda | 1 | 3 | 0 | 2 | 0 | 0 | 1 | 1 | 0 | 2 | 3 | 1 | 2 | 0 | 3 | 0 | 4 | 1 | 0 | 1 |
| B. aphidicola (S.g.) | 1 | 1 | 0 | 1 | 0 | 0 | 1 | 1 | 0 | 1 | 1 | 1 | 1 | 0 | 1 | 0 | 1 | 1 | 0 | 1 |
| B. aphidicola (B.p.) | 1 | 1 | 0 | 1 | 0 | 0 | 1 | 1 | 0 | 1 | 1 | 1 | 1 | 0 | 1 | 0 | 1 | 1 | 0 | 1 |
| B. aphidicola (A.p.) | 1 | 1 | 0 | 1 | 0 | 0 | 1 | 1 | 0 | 1 | 1 | 1 | 1 | 0 | 1 | 0 | 1 | 1 | 0 | 1 |
| E. coli CFT073 | 2 | 2 | 0 | 5 | 0 | 1 | 1 | 3 | 1 | 4 | 3 | 1 | 2 | 2 | 6 | 0 | 4 | 1 | 1 | 1 |
| E. coli K12 | 2 | 3 | 0 | 4 | 0 | 0 | 1 | 1 | 1 | 4 | 3 | 1 | 2 | 2 | 4 | 0 | 4 | 1 | 1 | 1 |
| E. coli O157:H7 EDL933 | 1 | 3 | 0 | 4 | 0 | 3 | 1 | 8 | 1 | 4 | 3 | 1 | 2 | 2 | 4 | 0 | 4 | 1 | 1 | 1 |
| E. coli O157:H7 | 2 | 3 | 0 | 4 | 0 | 3 | 1 | 8 | 1 | 4 | 3 | 1 | 2 | 2 | 4 | 0 | 4 | 1 | 1 | 1 |
| S. enterica | 2 | 3 | 0 | 4 | 0 | 0 | 1 | 3 | 1 | 4 | 3 | 1 | 2 | 1 | 4 | 0 | 3 | 1 | 1 | 1 |
| S. typhimurum | 2 | 3 | 0 | 4 | 0 | 0 | 1 | 1 | 1 | 4 | 3 | 1 | 2 | 2 | 2 | 0 | 3 | 1 | 1 | 1 |
| S. flexneri | 2 | 2 | 0 | 4 | 0 | 0 | 1 | 5 | 1 | 3 | 3 | 1 | 2 | 2 | 6 | 0 | 5 | 5 | 1 | 1 |
| Y.pestis C092 | 2 | 3 | 0 | 2 | 0 | 0 | 1 | 1 | 1 | 3 | 3 | 1 | 1 | 1 | 3 | 0 | 2 | 1 | 1 | 1 |
| Y. pestis KIM | 2 | 2 | 0 | 3 | 0 | 0 | 1 | 1 | 1 | 3 | 3 | 1 | 2 | 1 | 5 | 0 | 3 | 1 | 1 | 1 |
| | 52 | 101 | 9 | 91 | 3 | 10 | 33 | 60 | 29 | 82 | 94 | 47 | 66 | 23 | 103 | 8 | 110 | 46 | 21 | 43 |
| | # | # | | # | | | | # | | # | # | # | # | | # | | # | # | | # |

| | Ile | Leu-CUN | | | Leu-UUR | | Lys | | Met | Phe | Pro | | | Ser-UCN | | | Ser-AGY |
|---|---|---|---|---|---|---|---|---|---|---|---|---|---|---|---|---|---|
| | GAT | GAG | TAG | CAG | TAA | CAA | TTT | CTT | CAT | GAA | GGG | TGG | CGG | GGA | TGA | CGA | GCT |
| C. jejuni | 3 | 1 | 1 | 0 | 1 | 1 | 2 | 0 | 3 | 1 | 0 | 1 | 0 | 1 | 1 | 0 | 1 |
| H. pylori 26695 | 1 | 1 | 1 | 0 | 1 | 1 | 1 | 0 | 3 | 1 | 1 | 1 | 0 | 1 | 1 | 0 | 1 |
| H. pylori J99 | 1 | 1 | 1 | 0 | 1 | 1 | 1 | 0 | 3 | 1 | 1 | 1 | 0 | 1 | 1 | 0 | 1 |
| | | | | | | | | | | | | | | | | | |
| A. tumefaciens | 4 | 1 | 1 | 1 | 1 | 1 | 1 | 1 | 6 | 1 | 1 | 1 | 1 | 1 | 1 | 1 | 1 |
| S. meliloti | 3 | 1 | 1 | 1 | 1 | 1 | 1 | 0 | 5 | 1 | 1 | 1 | 1 | 1 | 1 | 1 | 1 |
| B. melitensis | 3 | 2 | 1 | 1 | 1 | 1 | 1 | 1 | 5 | 1 | 1 | 1 | 1 | 1 | 1 | 1 | 1 |
| B. sui | 3 | 2 | 1 | 1 | 1 | 1 | 1 | 1 | 5 | 1 | 1 | 1 | 1 | 1 | 1 | 1 | 1 |
| R. conorii | 1 | 1 | 1 | 0 | 1 | 0 | 1 | 0 | 3 | 2 | 0 | 1 | 0 | 1 | 1 | 0 | 1 |
| R. prowazekii | 1 | 1 | 1 | 0 | 1 | 0 | 1 | 0 | 3 | 2 | 0 | 1 | 0 | 1 | 1 | 0 | 1 |
| C. crescentus | 2 | 1 | 1 | 1 | 1 | 1 | 1 | 1 | 5 | 1 | 1 | 1 | 1 | 1 | 1 | 1 | 1 |
| | | | | | | | | | | | | | | | | | |
| R. solanacearum | 4 | 1 | 1 | 2 | 1 | 1 | 1 | 1 | 4 | 1 | 1 | 1 | 1 | 1 | 1 | 1 | 1 |
| N. meningitidis MC58 | 4 | 0 | 1 | 2 | 1 | 1 | 2 | 0 | 3 | 1 | 1 | 1 | 1 | 1 | 1 | 1 | 1 |
| N. meningitidis Z2491 | 3 | 1 | 1 | 2 | 1 | 0 | 2 | 0 | 2 | 1 | 1 | 1 | 1 | 1 | 1 | 1 | 1 |
| | | | | | | | | | | | | | | | | | |
| C. burnetii | 1 | 1 | 1 | 1 | 1 | 1 | 1 | 1 | 3 | 1 | 1 | 1 | 1 | 1 | 1 | 1 | 1 |
| X. fastidiosa | 2 | 1 | 1 | 1 | 1 | 1 | 2 | 1 | 3 | 1 | 1 | 1 | 1 | 1 | 1 | 1 | 1 |
| X. fastidiosa T. | 2 | 1 | 1 | 1 | 1 | 1 | 2 | 1 | 3 | 1 | 1 | 1 | 1 | 1 | 1 | 1 | 1 |
| X. axonopodis | 2 | 1 | 1 | 2 | 1 | 1 | 1 | 1 | 3 | 1 | 1 | 1 | 1 | 1 | 1 | 1 | 1 |
| X. campestris | 2 | 1 | 1 | 2 | 1 | 1 | 1 | 1 | 3 | 1 | 1 | 1 | 1 | 1 | 1 | 1 | 1 |
| P. aeruginosa | 4 | 2 | 1 | 2 | 1 | 1 | 2 | 0 | 4 | 1 | 1 | 1 | 1 | 1 | 1 | 1 | 1 |
| P. syringae | 5 | 1 | 1 | 2 | 1 | 1 | 2 | 0 | 5 | 1 | 1 | 2 | 0 | 1 | 1 | 1 | 1 |
| S. oneidensis | 3 | 1 | 2 | 2 | 3 | 1 | 8 | 0 | 8 | 3 | 1 | 3 | 0 | 1 | 3 | 0 | 2 |
| V. cholerae | 3 | 1 | 5 | 3 | 2 | 1 | 2 | 0 | 9 | 3 | 1 | 3 | 0 | 1 | 2 | 0 | 2 |
| V. parahaemolyticus | 2 | 2 | 10 | 0 | 3 | 1 | 4 | 0 | 11 | 4 | 0 | 3 | 0 | 1 | 4 | 0 | 1 |
| V. vulnificus | 3 | 2 | 7 | 0 | 2 | 1 | 4 | 0 | 9 | 4 | 1 | 3 | 0 | 1 | 4 | 0 | 2 |
| H. influenzae Rd | 3 | 1 | 1 | 0 | 2 | 1 | 3 | 1 | 4 | 1 | 0 | 2 | 0 | 1 | 2 | 0 | 1 |
| P. multicoda | 3 | 1 | 1 | 0 | 2 | 1 | 3 | 0 | 4 | 1 | 0 | 2 | 0 | 1 | 2 | 0 | 1 |
| B. aphidicola (S.g.) | 1 | 1 | 1 | 0 | 1 | 0 | 1 | 0 | 3 | 1 | 0 | 1 | 0 | 1 | 1 | 0 | 1 |
| B. aphidicola (B.p.) | 1 | 1 | 1 | 0 | 1 | 0 | 1 | 0 | 3 | 1 | 0 | 1 | 0 | 1 | 1 | 0 | 1 |
| B. aphidicola (A.p.) | 1 | 1 | 1 | 0 | 1 | 0 | 1 | 0 | 3 | 1 | 0 | 1 | 0 | 1 | 1 | 0 | 1 |
| E. coli CFT073 | 1 | 1 | 1 | 4 | 1 | 1 | 6 | 0 | 8 | 2 | 1 | 1 | 1 | 2 | 1 | 1 | 1 |
| E. coli K12 | 3 | 1 | 1 | 4 | 1 | 1 | 6 | 0 | 8 | 2 | 1 | 1 | 1 | 2 | 1 | 1 | 1 |
| E. coli O157:H7 EDL933 | 3 | 1 | 1 | 2 | 1 | 1 | 5 | 0 | 15 | 2 | 1 | 1 | 1 | 2 | 1 | 1 | 1 |
| E. coli O157:H7 | 3 | 1 | 1 | 3 | 1 | 1 | 5 | 0 | 15 | 2 | 1 | 2 | 1 | 2 | 1 | 1 | 1 |
| S. enterica | 3 | 1 | 1 | 4 | 1 | 1 | 4 | 0 | 6 | 2 | 1 | 1 | 1 | 2 | 1 | 1 | 1 |
| S. typhimurum | 3 | 1 | 1 | 4 | 1 | 1 | 5 | 0 | 6 | 2 | 1 | 1 | 1 | 2 | 1 | 1 | 1 |
| S. flexneri | 2 | 1 | 1 | 4 | 1 | 1 | 4 | 0 | 12 | 2 | 1 | 1 | 1 | 2 | 1 | 1 | 1 |
| Y.pestis C092 | 3 | 1 | 1 | 2 | 1 | 1 | 3 | 1 | 6 | 2 | 1 | 2 | 0 | 2 | 2 | 0 | 1 |
| Y. pestis KIM | 2 | 1 | 1 | 2 | 1 | 1 | 3 | 1 | 6 | 2 | 1 | 2 | 0 | 2 | 2 | 0 | 1 |
| | 94 | 42 | 58 | 56 | 46 | 32 | 95 | 13 | 210 | 59 | 29 | 52 | 21 | 47 | 51 | 22 | 41 |
| | # | | # | | # | | # | | # | # | | # | | # | # | | # |

| | Thr | | | Trp | Tyr | Val | | | Total tRNAs | 16S/23S | Genome length |
|---|---|---|---|---|---|---|---|---|---|---|---|
| | GGT | TGT | CGT | CCA | GTA | GAC | TAC | CAC | | | |
| C. jejuni | 1 | 1 | 0 | 1 | 1 | 1 | 2 | 0 | 42 | 1/1 | 1641481 bp |
| H. pylori 26695 | 1 | 1 | 0 | 1 | 1 | 1 | 1 | 0 | 36 | 2/2 | 1667867 bp |
| H. pylori J99 | 1 | 1 | 0 | 1 | 1 | 1 | 1 | 0 | 36 | 2/2 | 1643831 bp |
| | | | | | | | | | | | |
| A. tumefaciens | 1 | 1 | 1 | 1 | 1 | 1 | 1 | 0 | 53 | 1/1 | 2841581 bp |
| S. meliloti | 1 | 1 | 1 | 1 | 1 | 1 | 1 | 0 | 51 | 3/3 | 3654135 bp |
| B. melitensis | 1 | 1 | 1 | 1 | 1 | 1 | 1 | 1 | 54 | 3/3 | 3294931 bp |
| B. suis | 1 | 1 | 1 | 1 | 1 | 1 | 1 | 1 | 55 | 1/1 | 1207381 bp |
| R. conorii | 1 | 1 | 1 | 1 | 1 | 0 | 1 | 0 | 33 | 1/1 | 1268755 bp |
| R. prowazekii | 1 | 1 | 1 | 1 | 1 | 0 | 1 | 0 | 33 | 1/1 | 1111523 bp |
| C. crescentus | 1 | 1 | 1 | 1 | 1 | 1 | 1 | 1 | 51 | 2/2 | 4016947 bp |
| | | | | | | | | | | | |
| R. solanacearum | 1 | 1 | 1 | 1 | 1 | 1 | 1 | 1 | 57 | 3/3 | 3716413 bp |
| N. meningitidis MC58 | 1 | 2 | 1 | 1 | 1 | 1 | 2 | 0 | 57 | 4/4 | 2272351 bp |
| N. meningitidis Z2491 | 1 | 1 | 1 | 1 | 1 | 1 | 2 | 0 | 46 | 1/1 | 2184406 bp |
| | | | | | | | | | | | |
| C. burnetii | 1 | 1 | 1 | 1 | 1 | 1 | 1 | 0 | 42 | 1/1 | 1995275 bp |
| X. fastidiosa | 1 | 1 | 1 | 1 | 1 | 1 | 1 | 1 | 49 | 2/2 | 2679306 bp |
| X. fastidiosa T. | 1 | 1 | 1 | 1 | 1 | 1 | 1 | 1 | 49 | 2/2 | 2519802 bp |
| X. axonopodis | 1 | 1 | 1 | 1 | 1 | 1 | 1 | 1 | 54 | 2/2 | 5175554 bp |
| X. campestris | 1 | 1 | 1 | 1 | 1 | 1 | 1 | 1 | 53 | 2/2 | 5076188 bp |
| P. aeruginosa | 1 | 1 | 1 | 1 | 1 | 1 | 2 | 0 | 63 | 4/4 | 6264403 bp |
| P. syringae | 1 | 1 | 0 | 1 | 1 | 2 | 2 | 0 | 63 | 5/5 | 6397126 bp |
| S. oneidensis | 2 | 3 | 0 | 1 | 4 | 2 | 5 | 0 | 101 | 9/9 | 4969803 bp |
| V. cholerae | 2 | 4 | 0 | 1 | 5 | 2 | 2 | 0 | 98 | 8/8 | 2961149 bp |
| V. parahaemolyticus | 2 | 5 | 0 | 2 | 7 | 2 | 4 | 0 | 126 | 10/10 | 3288558 bp |
| V. vulnificus | 2 | 4 | 0 | 2 | 4 | 2 | 4 | 0 | 110 | 9/9 | 5260086 bp |
| H. influenzae Rd | 1 | 1 | 0 | 1 | 1 | 1 | 4 | 0 | 56 | 6/6 | 1830138 bp |
| P. multicoda | 1 | 1 | 0 | 1 | 1 | 1 | 4 | 0 | 56 | 1/1 | 2257487 bp |
| B. aphidicola (S.g.) | 1 | 1 | 0 | 1 | 1 | 1 | 1 | 0 | 32 | 1/1 | 641454 bp |
| B. aphidicola (B.p.) | 1 | 1 | 0 | 1 | 1 | 1 | 1 | 0 | 32 | 1/1 | 615980 bp |
| B. aphidicola (A.p.) | 1 | 1 | 0 | 1 | 1 | 1 | 1 | 0 | 32 | 1/1 | 640681 bp |
| E. coli CFT073 | 1 | 1 | 1 | 1 | 3 | 2 | 5 | 0 | 87 | 7/7 | 5231428 bp |
| E. coli K12 | 2 | 1 | 2 | 1 | 3 | 2 | 5 | 0 | 86 | 7/7 | 4639221 bp |
| E. coli O157:H7 EDL933 | 2 | 1 | 1 | 1 | 3 | 2 | 5 | 0 | 98 | 7/7 | 5528445 bp |
| E. coli O157:H7 | 2 | 1 | 1 | 1 | 3 | 2 | 5 | 0 | 101 | 7/7 | 5498450 bp |
| S. enterica | 2 | 1 | 1 | 1 | 1 | 2 | 3 | 0 | 77 | 7/7 | 4809037 bp |
| S. typhimurum | 2 | 1 | 1 | 1 | 3 | 2 | 5 | 0 | 79 | 7/7 | 4857432 bp |
| S. flexneri | 2 | 5 | 1 | 1 | 2 | 1 | 3 | 0 | 95 | 7/7 | 4607203 bp |
| Y.pestis C092 | 2 | 2 | 0 | 1 | 2 | 2 | 3 | 0 | 68 | 6/6 | 4653728 bp |
| Y. pestis KIM | 2 | 2 | 0 | 1 | 2 | 2 | 3 | 0 | 71 | 7/7 | 4600755 bp |
| | 50 | 57 | 24 | 40 | 67 | 49 | 88 | 8 | 2382 | | |
| | # | # | | # | # | | # | | | | |

Table 3 Lengths of the sequence alignments used.

| Alignment | Number of single sites | Number of pairs | Total length |
|---|---|---|---|
| Combined tRNA | 958 | 621 | 2190 |
| 16S (38 species) | 654 | 402 | 1458 |
| 23S (38 species) | 1402 | 699 | 2800 |
| 16S (96 species) | 597 | 387 | 1371 |
| 23S (96 species) | 1176 | 583 | 2342 |
| 16S (α group) | 617 | 380 | 1377 |
| 23S (α group) | 1406 | 463 | 2332 |

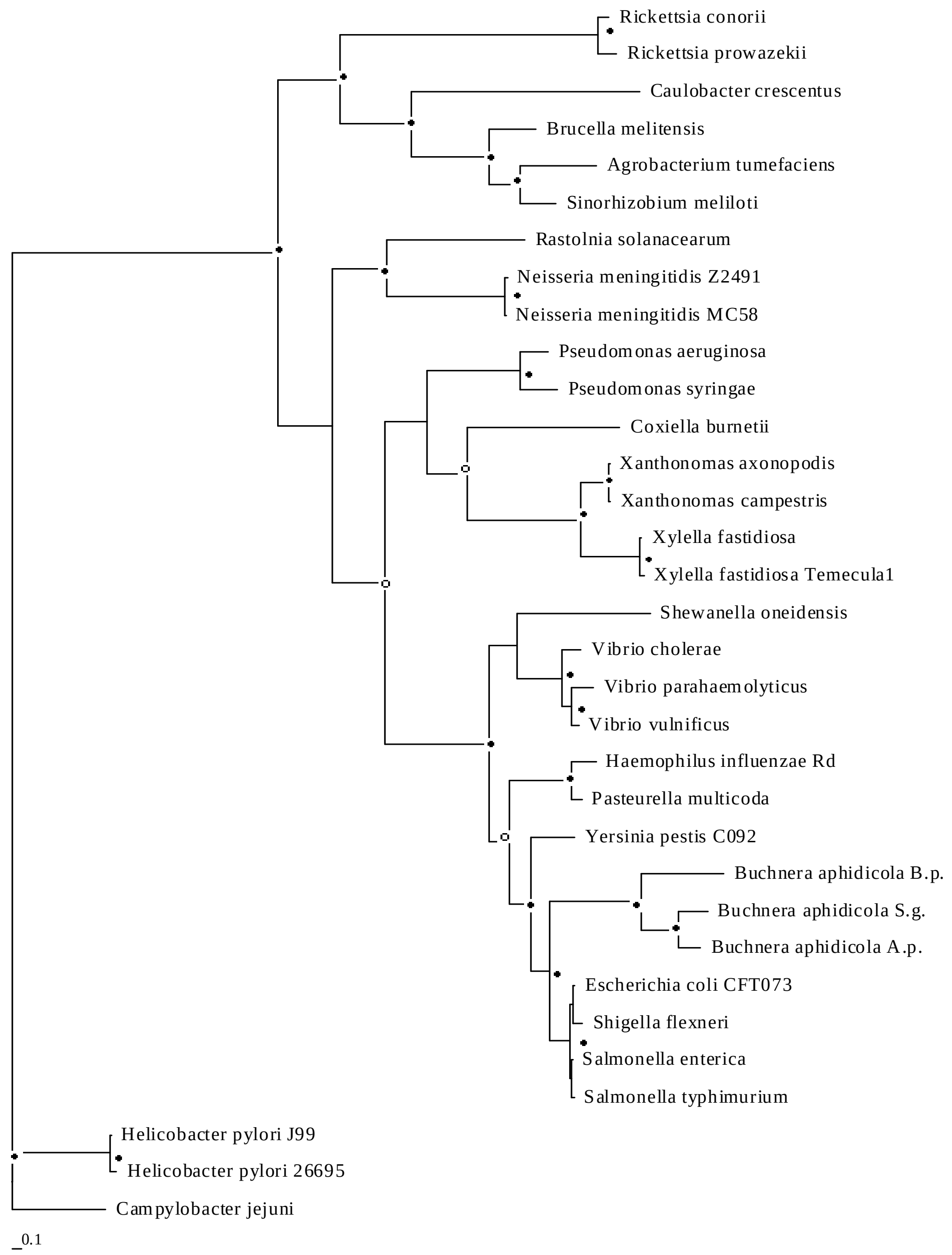

Figure 1 - Consensus phylogeny of the 29 concatenated tRNAs for the 38 species set. Branch lengths are the maximum likelihood values for the consensus tree topology. Nodes supported with 100% posterior probability in the Bayesian analysis are marked ● and those with posterior probability □90% are marked ○. The same notation is used in figures 2 – 7.

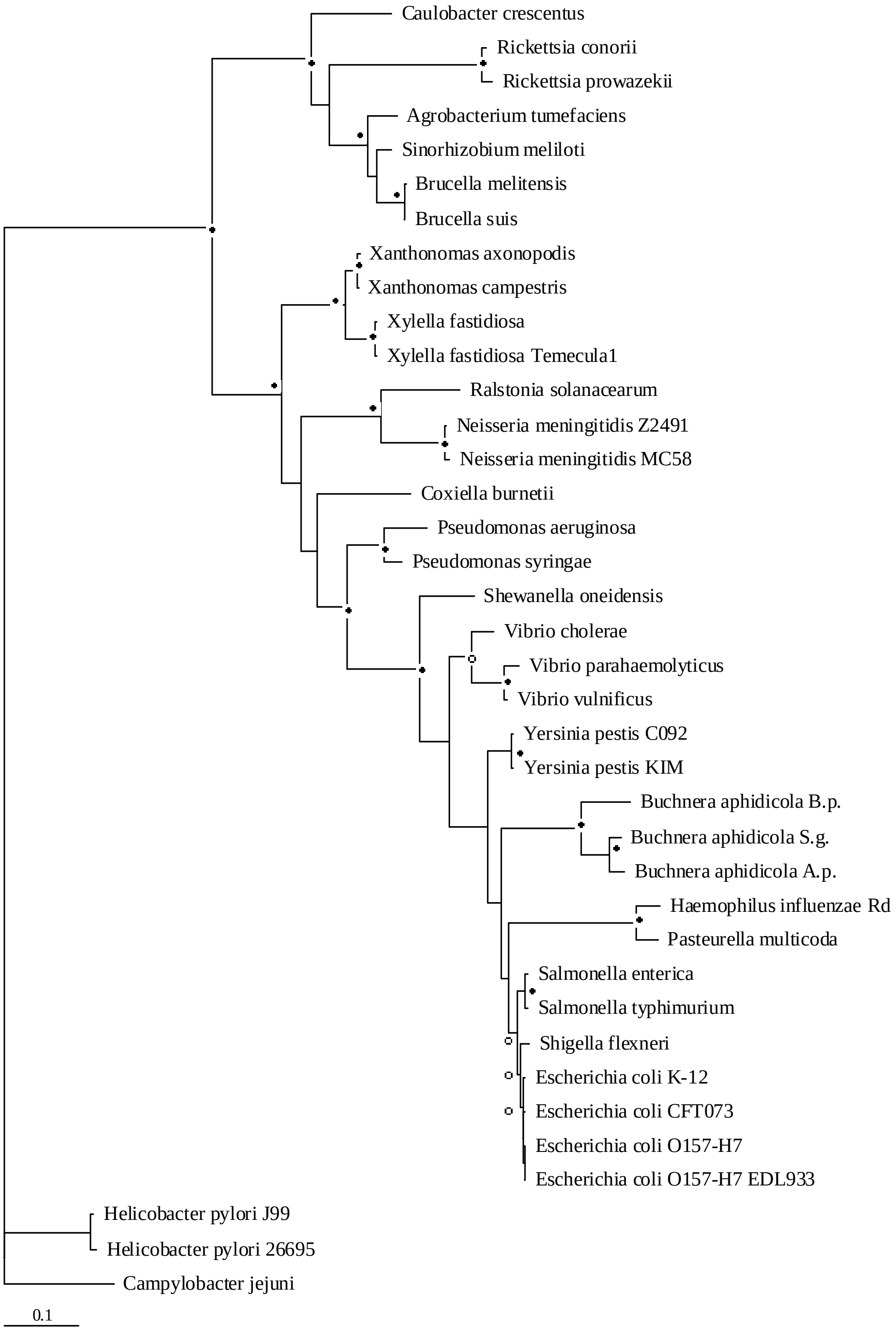

Figure 2 – Consensus phylogeny of 16S rRNA for the 38 species set.

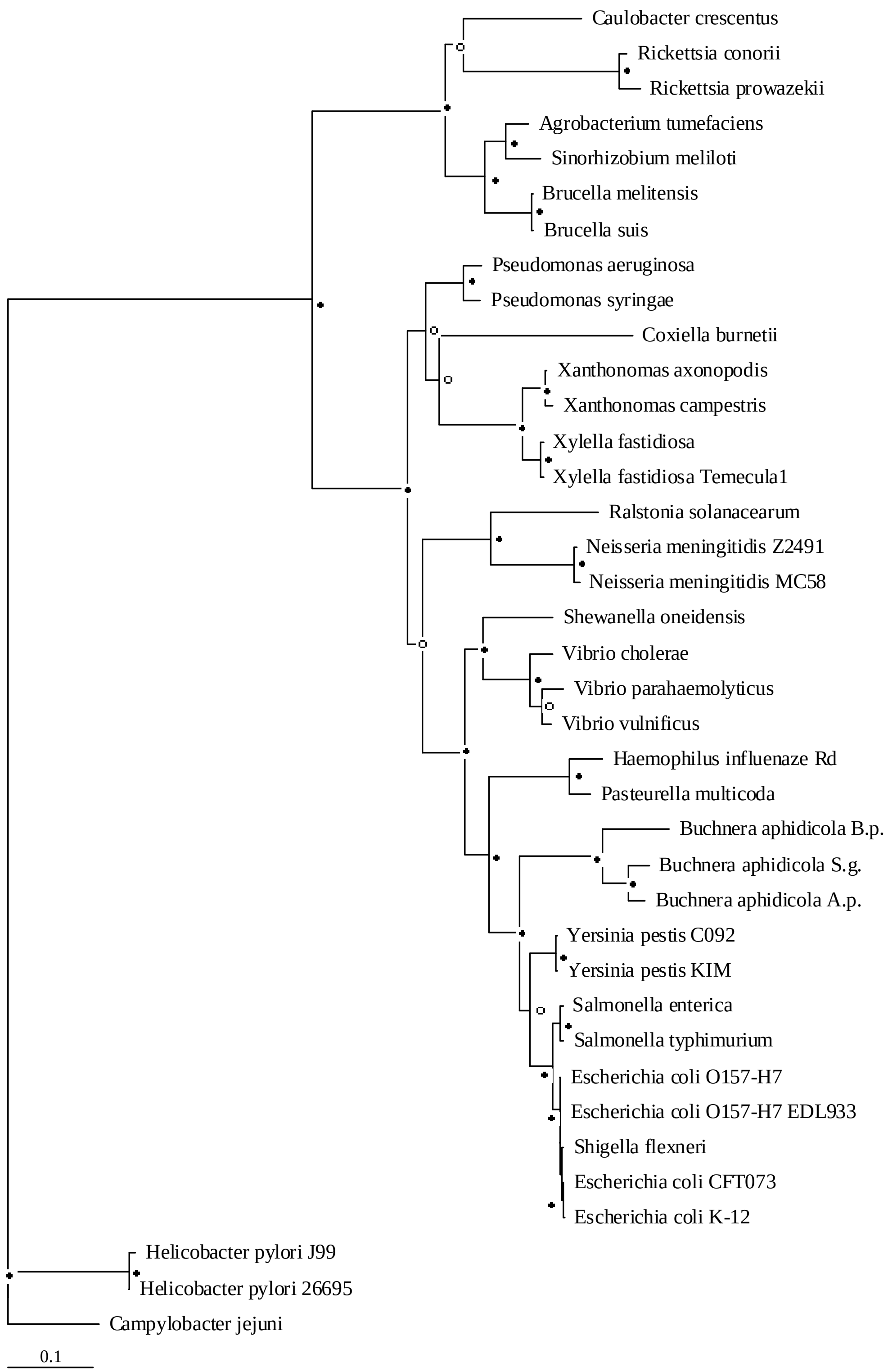

Figure 3 – Consensus phylogeny of 23S rRNA for the 38 species set.

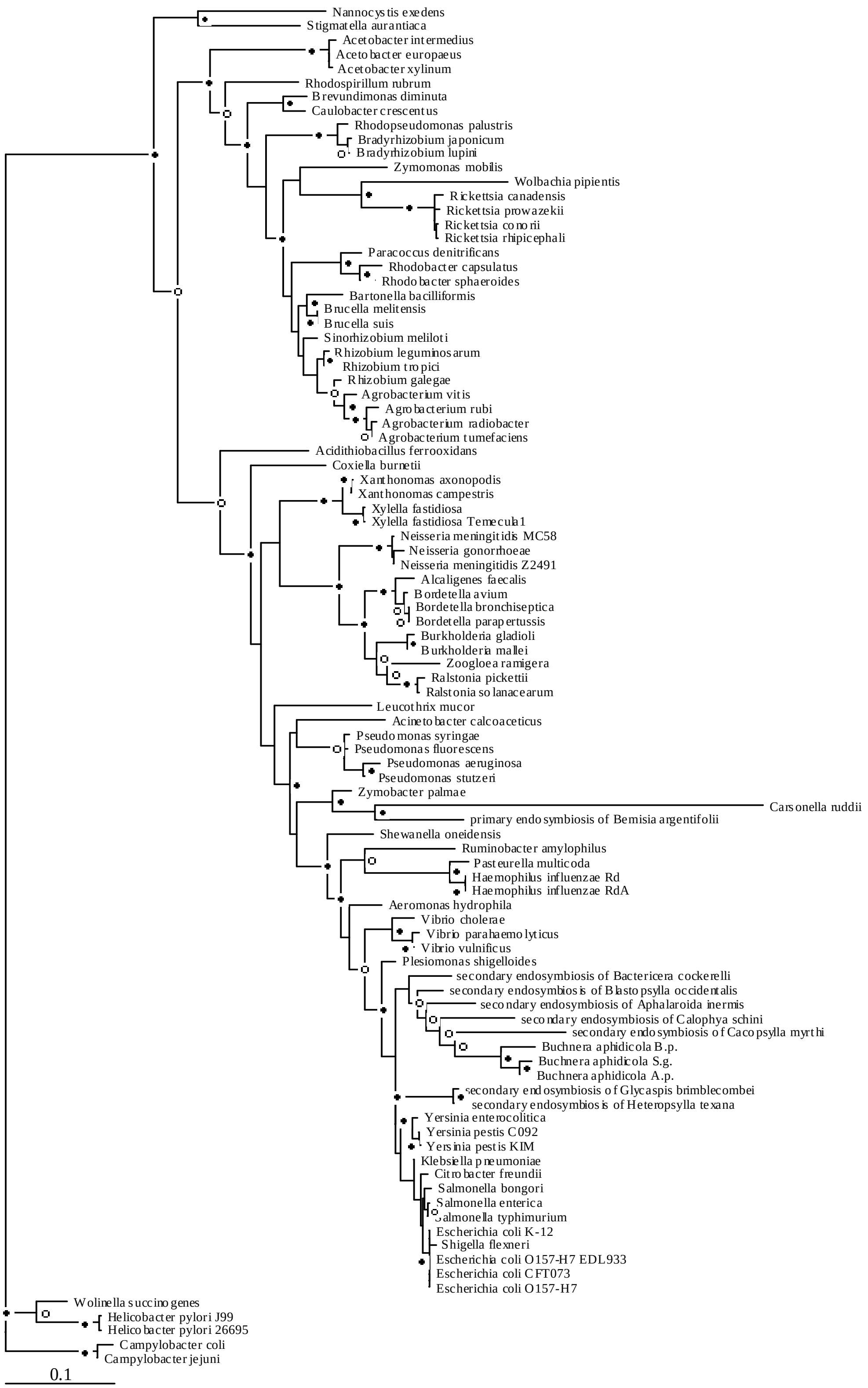

Figure 4 – Consensus phylogeny of 16S rRNA for the 96 species set.

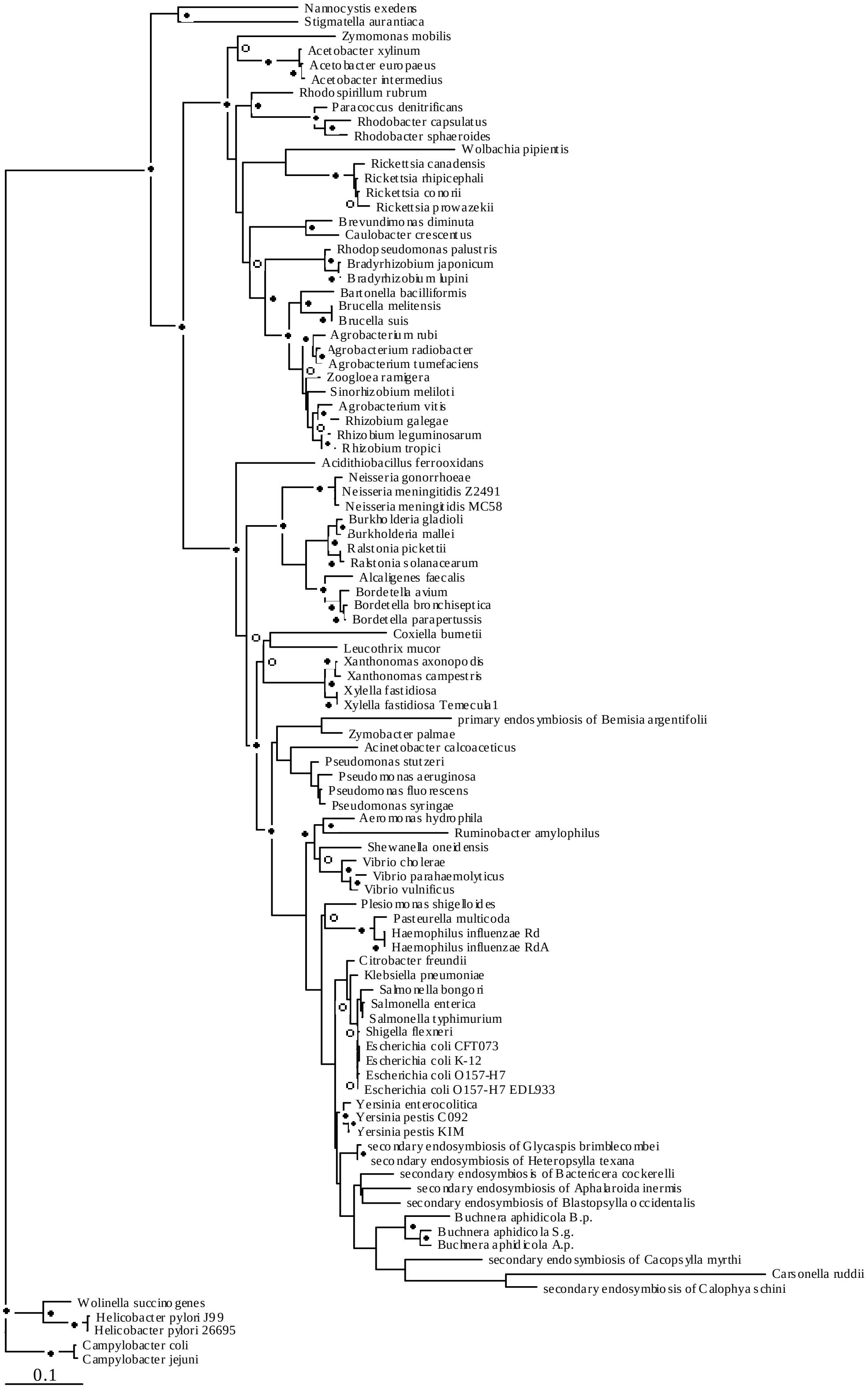

Figure 5 – Consensus phylogeny of 23S rRNA for the 96 species set.

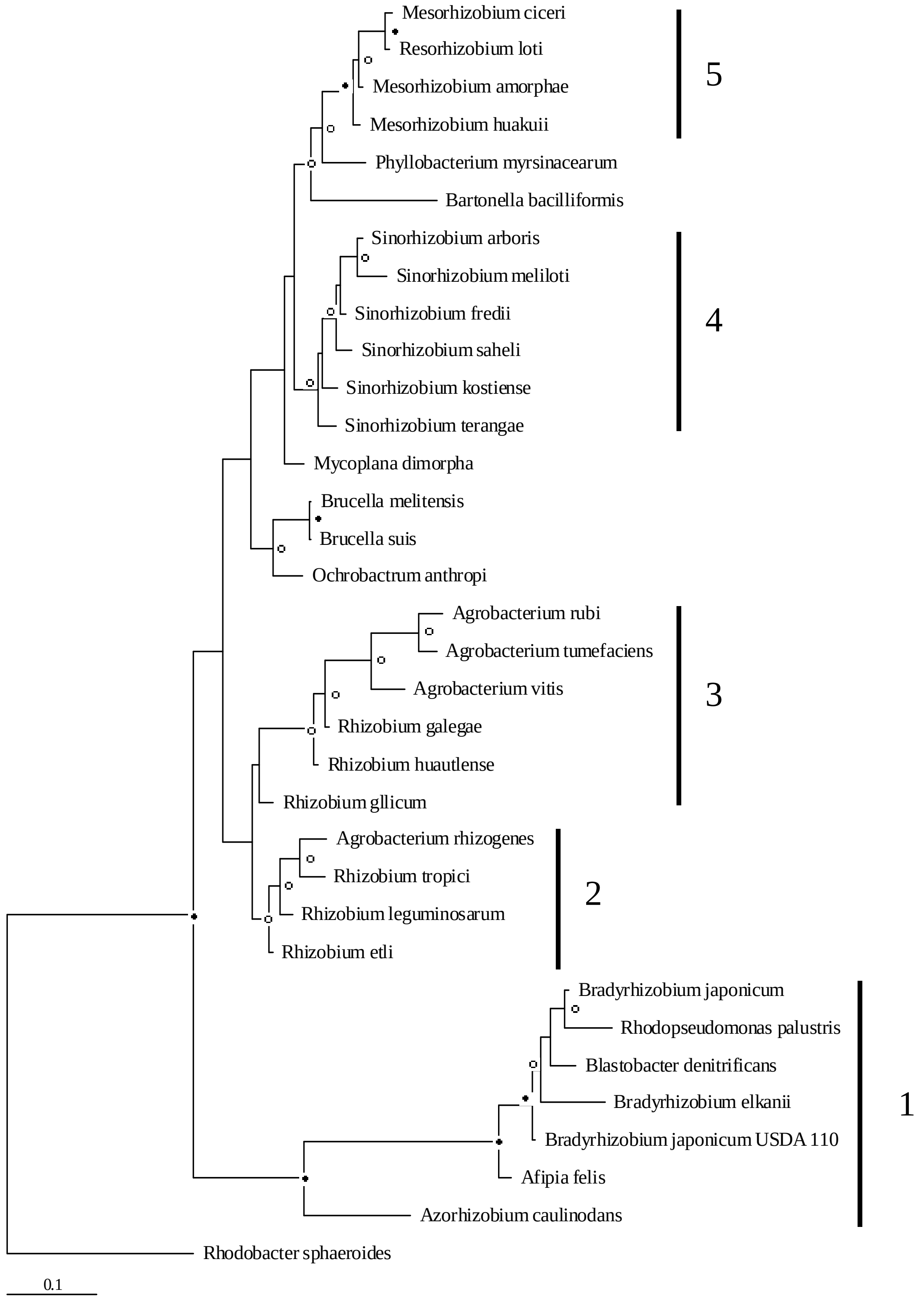

Figure 6 – Consensus phylogeny of 16S rRNA for the Rhizobiales.

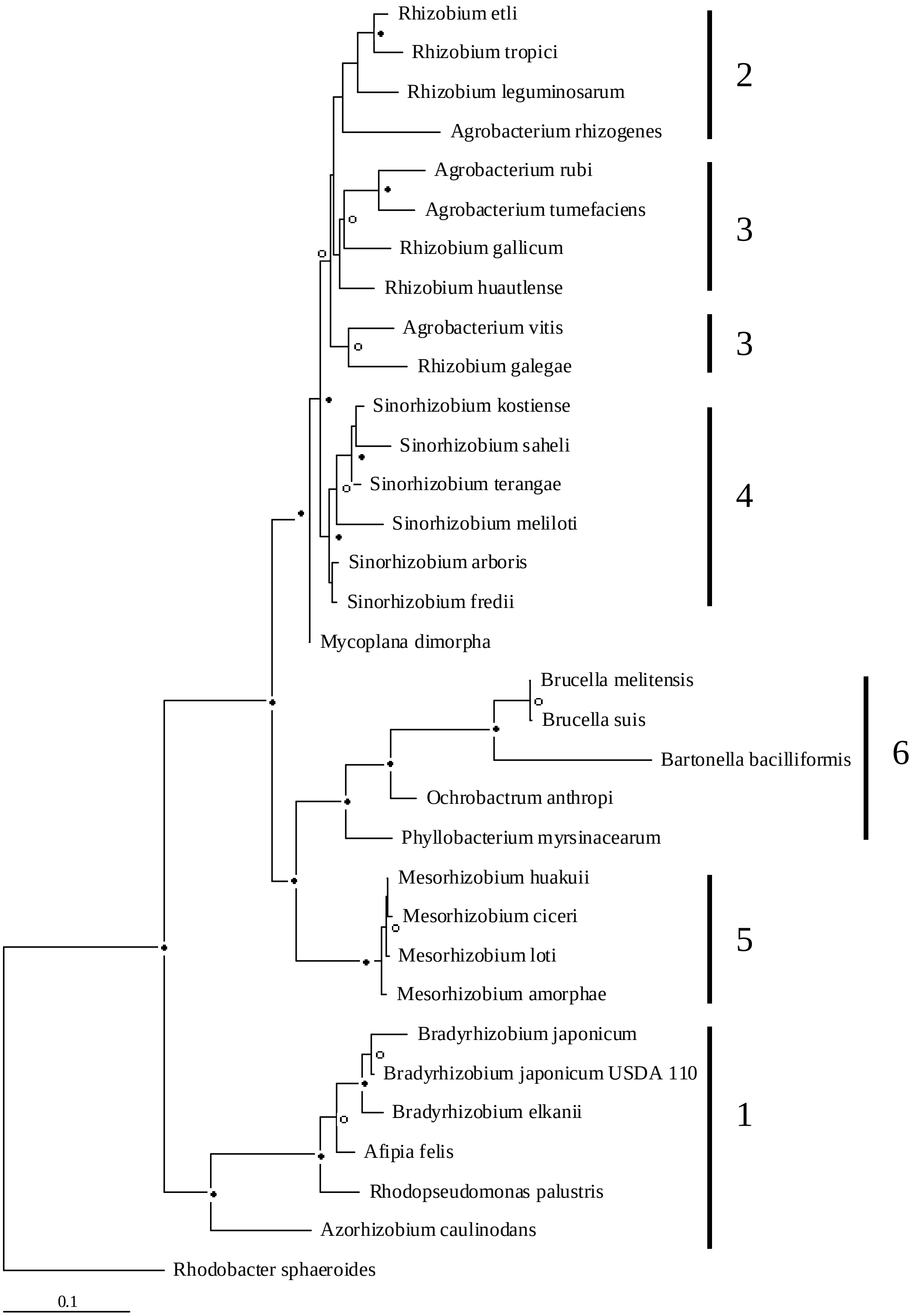

Figure 7 – Consensus phylogeny of 23S rRNA for the Rhizobiales.